\documentclass[a4paper]{jpconf}
\usepackage{graphicx}
\begin{document}
\title{The CMS Silicon Strip Tracker}

\author{Paolo Azzurri\footnote{on behalf of the CMS Tracker Collaboration}}

\address{Scuola Normale Superiore, piazza dei Cavalieri 7, 56100 Pisa, Italy}

\ead{paolo.azzurri@cern.ch}

\begin{abstract}
With over 200 square meters of sensitive Silicon and almost 10 million readout channels, 
the Silicon Strip Tracker of the CMS experiment at the LHC 
will be the largest Silicon strip detector ever built.
The design, construction and expected performance of the CMS Tracker is
reviewed in the following.
\end{abstract}
\section{The CMS experiment at the LHC}
The Compact Muon Solenoid (CMS) experiment is scheduled to start data taking in summer 2007 
at the Large Hadron Collider (LHC) at CERN where 7~TeV proton beams will collide head on 
at a center-of-mass energy of 14~TeV. The LHC is projected to reach a peak luminosity
of 10$^{34}$~cm$^{-2}$~sec$^{-1}$.
The LHC will operate with a bunch crossing rate of 40~MHz (25~ns) and at peak luminosity about 20 
interactions are expected per crossing, producing on average 2000 charged particles per crossing.
Severe radiation conditions are expected, corresponding to a 
flux of 10$^{18}$ hadrons per year from interaction points.
\begin{figure}[h]
\includegraphics[width=24pc]{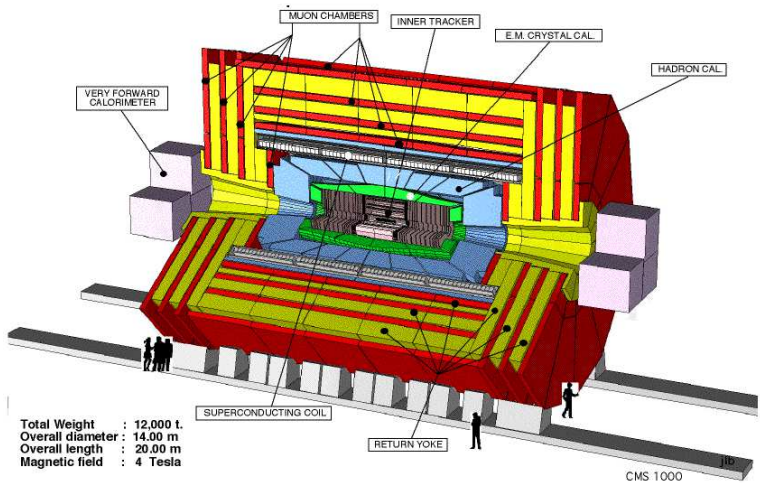}\hspace{2pc}%
\begin{minipage}[b]{12pc}\caption{\label{fig:cms}
A three-dimensional view of the CMS experiment and it's various subdetectors.
The Silicon Strip Tracker will be placed in the inner part of the experiment, around the 
beam line and the interaction point, immersed in the 4~Tesla solenoidal magnetic field.   
}
\end{minipage}
\end{figure}
 
The complete CMS experiment will be a cylinder 20~m long and 14~m in diameter, 
with 12,000 tons of total weight. CMS will be composed of an external 
high precision muon spectrometer in the return yoke 
of a 4~Tesla superconducting solenoidal coil cylinder 12.5~m long and 6~m in diameter. 
The coil will contain  a sampling brass hadron calorimeter, 
a lead-tungstate scintillating electromagnetic calorimeter, 
 and, closest to the beam line, an internal Silicon Tracker, as shown in figure~\ref{fig:cms}.

\section{The CMS Inner Tracker}
\begin{figure}[htb]
\begin{center}
\includegraphics[width=15cm]{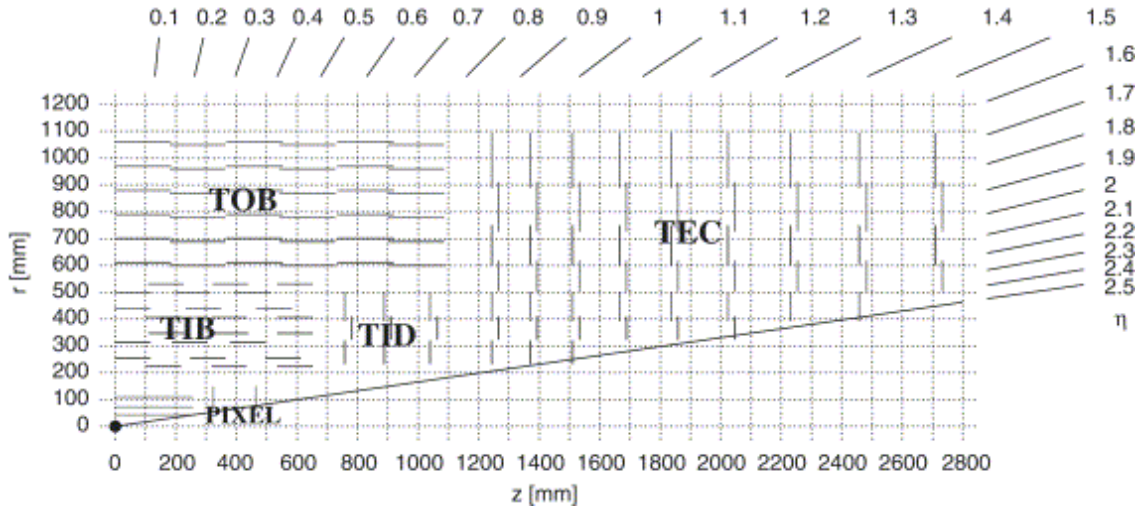}
\caption{\label{fig:sst}
Projected view of one quarter of the CMS tracker layout in the $r-z$ plane, showing the pseudorapidity
coverage. Segments represent detector modules, lighter ones are single sided and darker ones
double-sided.}
\end{center}
\end{figure}

The CMS Silicon tracker~\cite{cmst} 
is composed of different substructures.
Closest to the Interaction Point (IP) is a Silicon Pixel detector,
with about 66 million 100$\times$150~$\mu$m$^2$ pixels 
arranged at distance of 4 to 11~cm from the beam line 
on a cylindrical barrel and end-caps structure with total length of 92~cm,
this detector will not be further described in this paper.
The Silicon Strip detectors are divided in the inner barrel part (TIB),
the inner disks (TID), the outer barrel (TOB) and outer end-caps (TEC).
The layout of the Tracker substructures is sketched in figure~\ref{fig:sst}.

As it can be seen in figure~\ref{fig:sst}, the TIB and TOB  systems are composed respectively 
of four and six concentric layer barrel shell structures.
The TID system is made of three disk structures on each side, each divided in 
three concentric rings, while the TEC is made of nine
disk structures on each side, each made of four to seven rings. 
The whole inner tracker will be housed in a cylindrical support structure
with a diameter of 2.4~m and a total length 
of 5.6~m. An active thermal screen will keep the tracker volume at at temperature of -10$^\circ$C
and at 30\% relative humidity, to avoid the reverse annealing of the silicon sensors, and to 
protect the silicon detectors from the increased leakage current 
coming from radiation damage. A cooling system will extract the 60~kW power that the 
front-end electronics dissipate.

\section{Tracker Modules}
\begin{figure}[htb]
\begin{center}
\includegraphics[width=15cm]{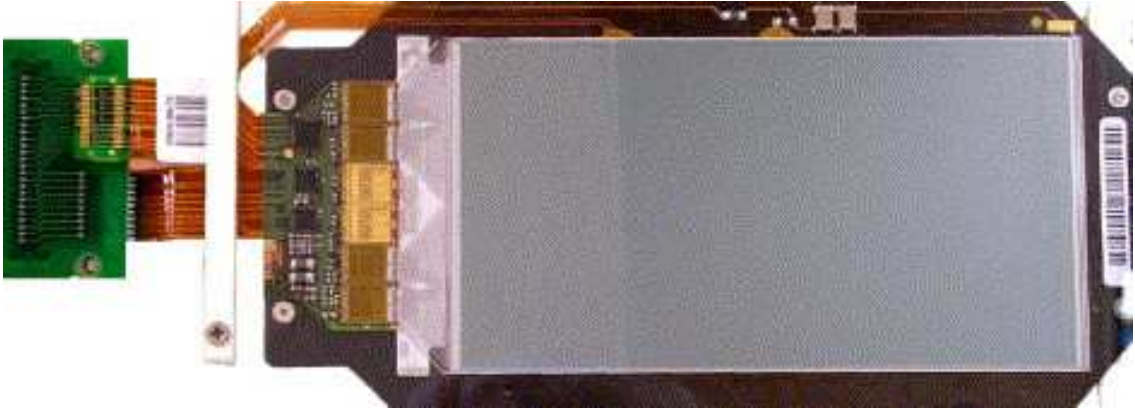}
\caption{\label{fig:mod}
A module for the CMS tracker inner barrel (TIB). The external black structure 
is the carbon fiber frame. The larger rectangular central grey area is the silicon sensor,
surrounded by the kapton circuit for insulation and bias. At the left edge of the sensor,
in light grey, is the glass pitch adapter, bringing the signal to the front-end hybrid
and chips, on the far left of the frame, and to the cable out of the module on the left.}
\end{center}
\end{figure}

The Tracker will be composed of 15,148 detector modules distributed among the four 
sub-systems (TIB, TID, TOB, TEC) described above. 
Each module has one or two silicon sensors, for a total of 24,244 sensors. 
Modules are supported by a carbon-fiber or graphite frame, with a kapton 
layer to isolate the silicon and provide the electrical connections to the sensor backplane. 
The readout and auxiliary chips are housed on a ceramic multiplayer hybrid, and 
a glass pitch adapter between the hybrid and the sensor, brings the signals 
from the sensor strips to the readout input pads. 
An example of a TIB detector module is shown in figure~\ref{fig:mod}.

The front-end hybrid hosts four or six APV25 readout chips. 
The APV25 chip has 128 amplifying channels and is designed 
in 0.25~$\mu$m CMOS technology to be radiation hard 
with low noise and a fast signal readout~\cite{apvc}.
The signal shaping with a de-convolution filter has a shaping time of 25~ns.
Further a pipeline buffer of 192 columns can store LHC bunch crossings over 4.8~$\mu$s,
to allow a decision from the CMS first level trigger system.

In each module large number of micro-bond wire connections are necessary to bring signals between
(i) the sensors, if two are present, (ii) the sensor and the pitch adapter, and 
(iii) the pitch adapter and the readout hybrid. 
For all modules, a total of approximately 25 million wire bonds are necessary,
and are made with the help of programmable automatic micro-bonding stations.  
 
\begin{figure}[htb]
\begin{center}
\includegraphics[width=12cm]{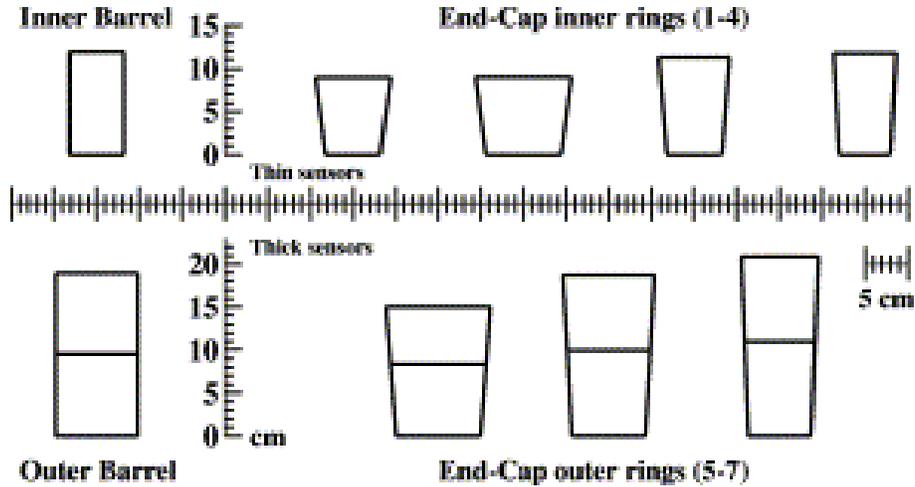}
\caption{\label{fig:modt} Shapes and dimensions of the CMS tracker 
modules mounted on the Inner Barrel (TIB, upper left), Outer Barrel (lower left)
and End-Caps inner (upper right) and outer (lower right) rings .}
\end{center}
\end{figure}

 The CMS tracker modules come in a variety of shapes and dimensions. The 
 outer modules of the TOB structure and of the three outer TEC rings, hold two sensors,
 all the other TIB, TID and four inner TEC rings, have a single sensor. 
 As for the modules shapes, the TIB and TOB barrel modules are 
 rectangular, while the TID and TEC disc modules have a wedge shape, in
 order to form rings. A sketch with the different shapes and dimensions of
 the tracker modules is shown in figure~\ref{fig:modt}
 
Roughly half of the modules in the tracker layout are in fact 
double-sided modules, made of two independent single-sided modules
glued together back-to-back with a relative rotation of 100~mrad 
respect to each other.
This allows a determination of the ionization in  the $z$ 
coordinate in the barrel modules, and in the $r$ coordinate 
in the disks, i.e. the determination of a full space point where the 
double sided modules are present.  
These double-sided modules are 
mounted in the first two layers of the TIB and TOB, 
in the first two TID rings,
and in rings 1, 2 and 5 of the TEC structure.

\section{Tracker Sensors}
The design of the CMS tracker sensors is rather simple, and has been
studied in collaboration with industry to ease the mass production needed
to build such a large silicon system. This is why only single-sided sensors
are produced. Also the 15 different sensor types finally designed for the 
tracker make use of the maximum area available on the 6~inch wafers 
commonly used in sensor production lines. 

The choice of the material is of n-type silicon with 512 or 768 p$^+$ single sided strips,
of resistivity in the 1.25-7.5~K$\Omega$~cm range, with thickness of 320~$\mu$m
or 500~$\mu$m, and  $<$100$>$ crystal orientation.
Design choices have been driven by many factors, and most of all radiation 
hardness.

Silicon is inherently radiation hard and the most important macroscopic effects 
of irradiation are the increase of the leakage current, that is linear with the 
radiation fluence, and the conversion of the n-type bulk into p-type bulk
with the increase of the carrier concentration.
As a consequence the bias voltage needed for depletion increases with the 
radiation damage, and this can lead to the detector current breakdown. 
For these reasons the inner parts of the tracker 
(TIB, TID and the four innermost TEC rings)
have been instrumented with sensors of lower resistivity (1.25-3.25~K$\Omega$~cm)
and standard 320~$\mu$m thickness, while the outer parts of the tracker 
have higher resistivity (3.5-7.5~K$\Omega$~cm)
and thicker 500~$\mu$m sensor wafers.

The lower resistivity of the inner sensors will require higher operational 
voltages at startup (300V) but this will give more margin for the 
bulk type inversion caused by 
the damage from higher radiation levels closer to the interaction point.
In this way even at the end of 10 years of operation, even the sensors that
are most exposed to irradiation will require maximum operational voltages, after the
bulk inversion, similar to the startup ones. 
%
The choice of the lattice orientation $<$100$>$ is also driven by radiation hardness 
considerations as it minimizes the surface damage and thus 
the increase of inter-strip capacitance  after irradiation~\cite{brai}.

\begin{figure}[htb]
\begin{center}
\includegraphics[width=12cm]{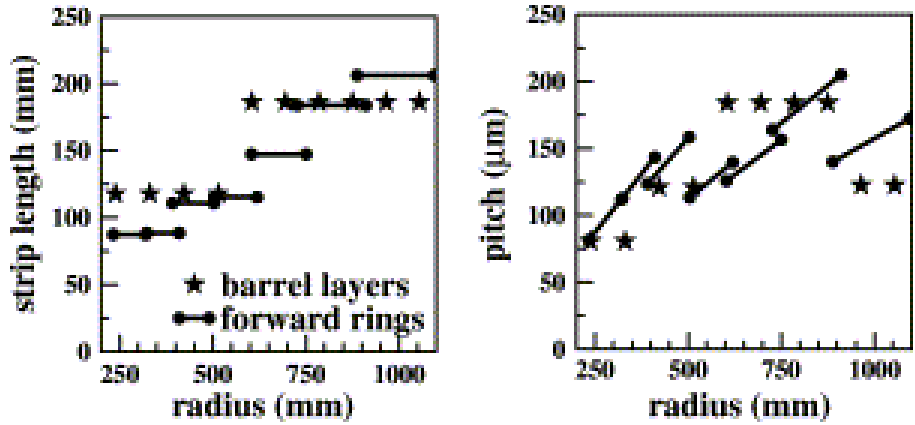}
\caption{\label{fig:mods} Silicon sensors strips length and pitch as a function
of the module radius to the beam line. Markers are barrel sensors, lines are end-cap
ring sensors.}
\end{center}
\end{figure}

The inter-strip distance (strip pitch) in different sensors varies from 80~$\mu$m
to 205~$\mu$m, but the ratio of the strip pitch to the strip width is 0.25 for all types.
The strips width, pitch and length are chosen in order to minimize inter-strip 
capacitance, optimize the resolution and occupancy, and assure high voltage
operational stability. The evolution of strip length and pitch with the radius of the 
detector position to the beam line is shown in figure~\ref{fig:mods}.
In the inner 320~$\mu$m thick sensors the strip lines cover a surface of 0.1~cm$^2$
per channel, in the outer 500~$\mu$m thick sensors strip lines cover 0.4~cm$^2$,
this leads to an expected occupancy around 1\% with the expected LHC track densities
at high luminosity.

\begin{figure}[h]
\includegraphics[width=18pc]{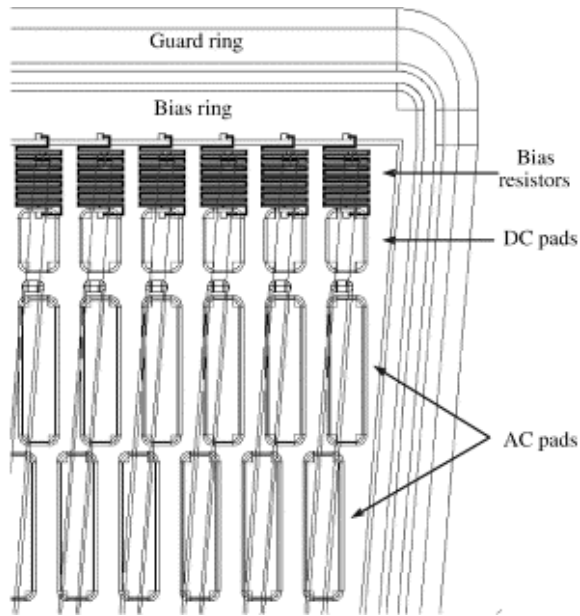}\hspace{2pc}%
\begin{minipage}[b]{18pc}\caption{\label{fig:sens}
View of a corner region of a wedge-type silicon strip sensor for the CMS tracker.
The bias ring runs around the sensor and brings the depletion voltage to the p$^+$
strips through the bias resistors. The floating guard ring runs around the bias ring and 
protects the sensor at high voltage operations. The series of DC pads are in
direct contact with the p$^+$ strip implants, while the two series of AC pads
are in contact with the superficial aluminum strips, and serve for the wire 
bonding connections to the readout electronics. }
\end{minipage}
\end{figure}

Strips are AC coupled to aluminum readout lines of 1.2~$\mu$m minimum thickness, 
and the coupling insulation is achieved with two layers of dielectric SiO$_2$
and Si$_3$N$_4$.
The high voltage operational stability of the module is enhanced by the 
metal overhanging of the aluminum readout strips over the p$^+$
implants, and by introducing a floating p$^+$ guard ring around the 
p$^+$ bias ring, to avoid high field near cut edges.
The bias ring is connected to the p$^+$ strips with an array of polysilicon 
resistors, each 1.5$\pm$0.5~M$\Omega$. DC pads in direct contact with
the p$^+$ strips implants are available for testing, while two series of AC pads
are available for the wire micro-bonding of each strip to the readout electronics.
A sketch view of a typical corner of a CMS tracker silicon sensor is shown in 
figure~\ref{fig:sens}.

\section{Sensor production and quality assurance}
The production of the CMS silicon sensors has been awarded to two companies,
Hamamatsu Photonics (Hamamatsu, Japan) for thin (320~$\mu$m) sensors
and ST Microelectronics (Catania, Italy) for thick (500~$\mu$m) sensors.
The acceptance test for the quality assurance of produced sensors have been performed 
in four stages. The first stage of measurements is done at the manufacturing centers, 
while the following three are done 
at CMS centers that perform a full sensor {\it quality control}, a {\it process control} 
on test structures and {\it irradiation test} on a small fraction of sensors.

The sensor quality control, performed at  Karlsruhe, 
Rochester, Perugia, Pisa and Vienna, include 
(i) an optical inspection of possible sensor damages, 
(ii) a scan of the sensor capacitance and leakage
current as a function of the bias voltage up to 550~V, (iii) a measurement of strip parameters 
with a 400~V bias. 
The process controls, performed at Florence, Strasbourg and Vienna, are another set of
ten standard electrical measurements to be compared to the specifications and to
monitor the production. 
The irradiation test are performed at Louvain and Karlsruhe on small samples of sensors 
at -10$^\circ$C up to the final doses expected in CMS, and serve to confirm the 
expected radiation hardness of the detectors. 

Both the quality and process control test have revealed different problems in the sensor 
production and feedback has been given to the manufacturing companies to correct
the production faults. Just recently (2005) the full production of all fully qualified silicon sensors 
for the CMS tracker has been accomplished.

\section{Module production and quality assurance}
The assembly of the 15,232 modules of the tracker is performed in a semi-automatic way
by gantry stations. Such stations can localize specific markers with a pattern recognition 
program and, using pickup tools, can position and glue the module components with 
precisions better than 10~$\mu$m in positioning and better than 5~mrad in alignment. 
The six CMS assembly stations operate at a rate of 90 modules per day.

After the assembly stage, the modules are sent to the bonding and testing centers,
where the sensor strips are bonded to the front-end hybrid channels, and different
electrical test are performed~\cite{arct}. The test procedures include 
(i) an optical inspection of possible damages, 
(ii) a scan of the sensor leakage current with bias voltage up to 450~V,
(iii) pedestal, noise and pulse shape measurements for each channel 
in four different data acquisition modes, 
(iv) a LED illumination test of the sensor, to spot open or pinhole channels.
Channels with shorts or pinholes can cause readout problems; they are therefore 
disconnected and become open channels.
At the end of the test each channel can be flagged as bad because it is  
too noisy, open, dead or has other readout problems. 

Currently (December 2005) over 10,000 modules have been assembled and tested at 
CMS production centers, and about 3.5\% of produced modules do not pass the quality 
requirements.  
A module fails the requirements on the leakage current behavior if either it has a current
larger than 10~$\mu$A at 450~V, or if it shows a resistive/breakdown I-V curve at lower
voltages. About 1\% of produced modules fail these I-V requirements.
Another 0.5\% of produced modules have a number of bad channels greater that 2\% of the total
channels, and are classified bad for strips, while the total number of bad strips in good modules
is at the level of 0.1\%.
Problems with the front-end hybrid electronics cause another 0.5\% of produced modules 
to be classified as bad. The remaining bad modules (1-1.5\%) have other assembly
problems but a good fraction of these can probably be recovered.

After the standard full electrical tests modules go through a {\em long term} test where they are 
thermally cycled in the -20$^\circ$C to +25$^\circ$C range for 72 hours in a dry box and 
readout in operating conditions are simulated. Only 0.1\% of produced modules are seen 
to fail at this stage of testing. 
All the bad modules not passing the quality requirements are collected in four 
specialized repair centers for further diagnosis and possible recovery.
 
\section{Integration and test of larger structures}
\begin{figure}[h]
\includegraphics[width=22pc]{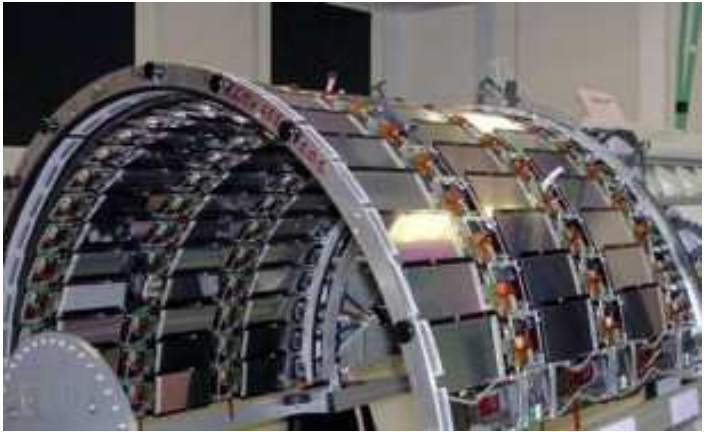}\hspace{2pc}%
\begin{minipage}[b]{14pc}\caption{\label{fig:lay3}
One half of the inner barrel layer three shell assembled and cabled at the INFN Pisa
integration facility. Single-sided silicon strip modules are mounted on both the internal
and external surfaces of the barrel.}
\end{minipage}
\end{figure}

As qualified modules are built they are also mounted on their final structures 
at the integration centers. At present hundreds of modules have been 
assembled on their supporting mechanics and subsystems, like the one 
pictured in figure~\ref{fig:lay3}. 
Coming out from the modules front-end hybrids, the data are converted into 
optical signals, travels in optical fibers to the back-end, where they are converted back
into electrical signals, digitized and processed. 

Substructures like the one in figure~\ref{fig:lay3} are also tested inside 
large cold rooms, where the functioning of the cooling, controls and data 
acquisition of large substructures can be tested in normal operational 
conditions. Starting at the beginning of 2006 the substructures assembled 
and tested at the integration centers will be transported to the CERN
Tracker Integration Facility (TIF), where finally the full tracker structure 
will be assembled and tested. 
The completed and tested silicon tracker will then be delivered at the CMS
experimental area at the end of 2006. 
In the meanwhile a test of smaller substructures of the tracker, built with modules 
that were not fully qualified, will also be tested inside the operating CMS magnet,
with parts of other CMS detectors, with the goal of triggering and measuring
cosmic rays.

\section{Tracker physics performances}
The CMS silicon strip tracker will be located inside the 4~Tesla solenoidal magnetic field 
and will provide measurements of 10 to 14 hits for charged tracks originating near the 
interaction point within a pseudorapidity acceptance of $|\eta|<$2.5.
Combinatorial Kalman filters are used to reconstruct charged tracks trajectories 
both with {\em inside-out} and {\em outside-in} seeding techniques~\cite{perf}.

\begin{figure}[htb]
\begin{center}
\includegraphics[width=5.2cm] {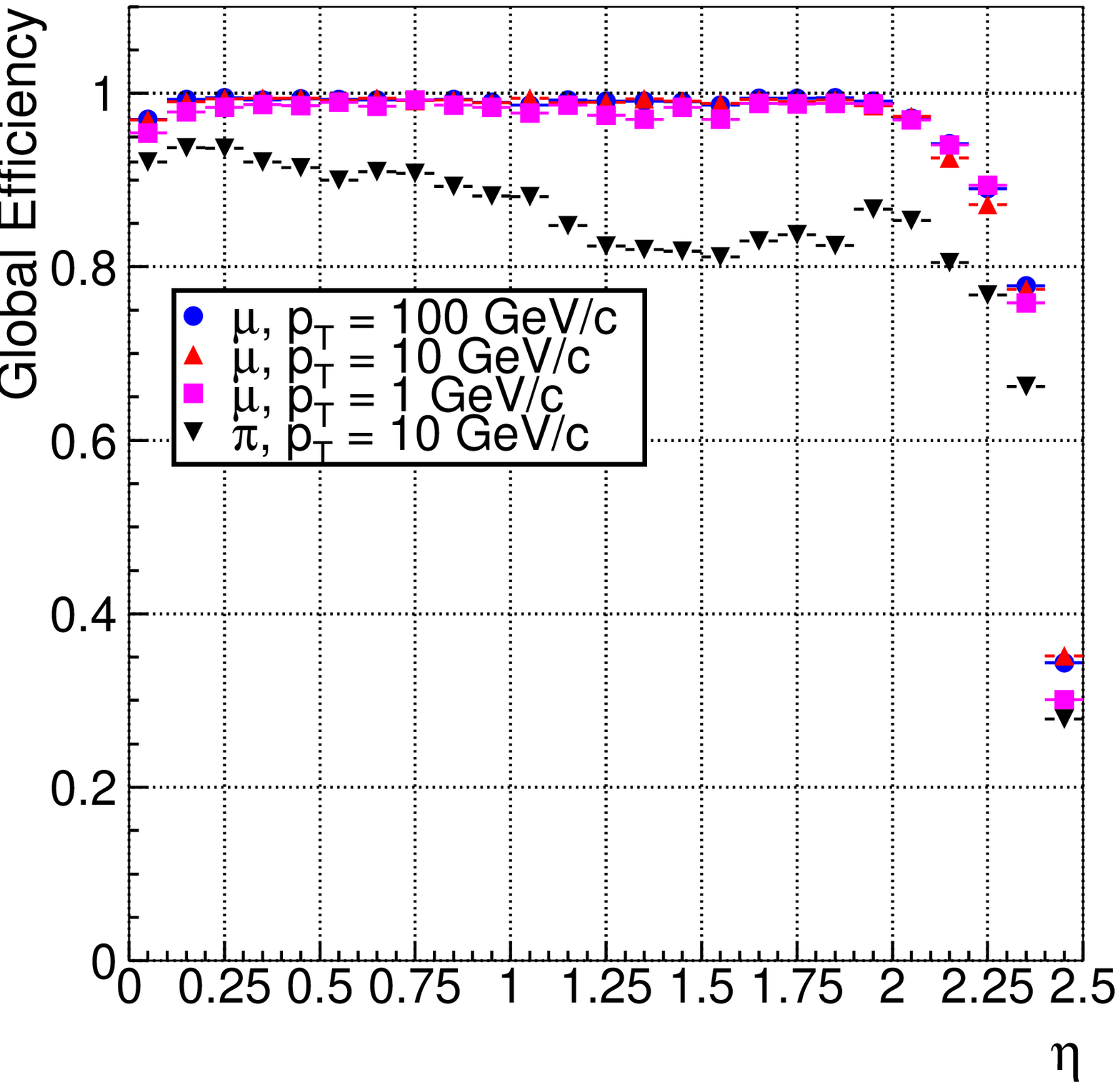}
\includegraphics[width=5.2cm] {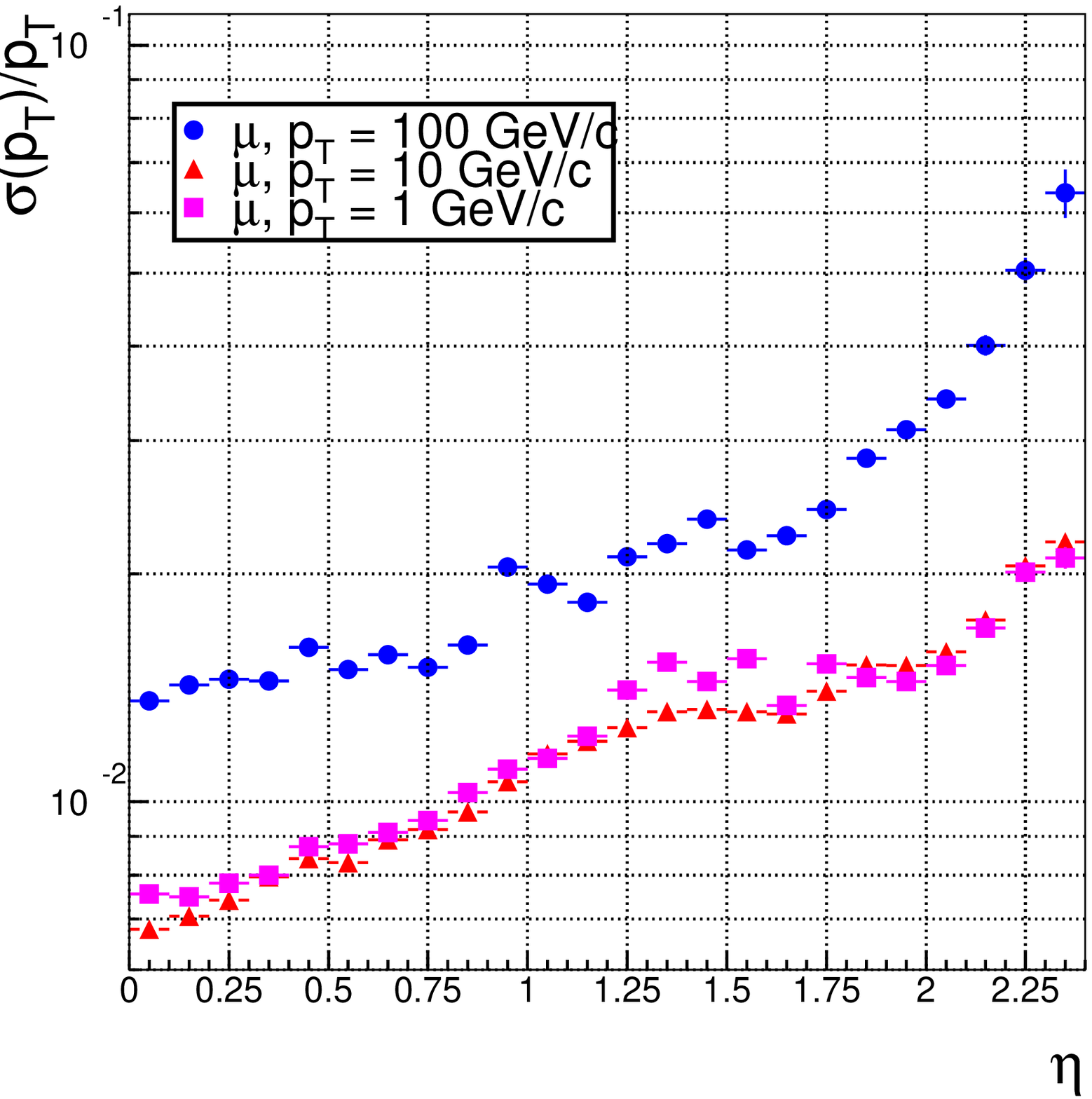}
\includegraphics[width=5.2cm] {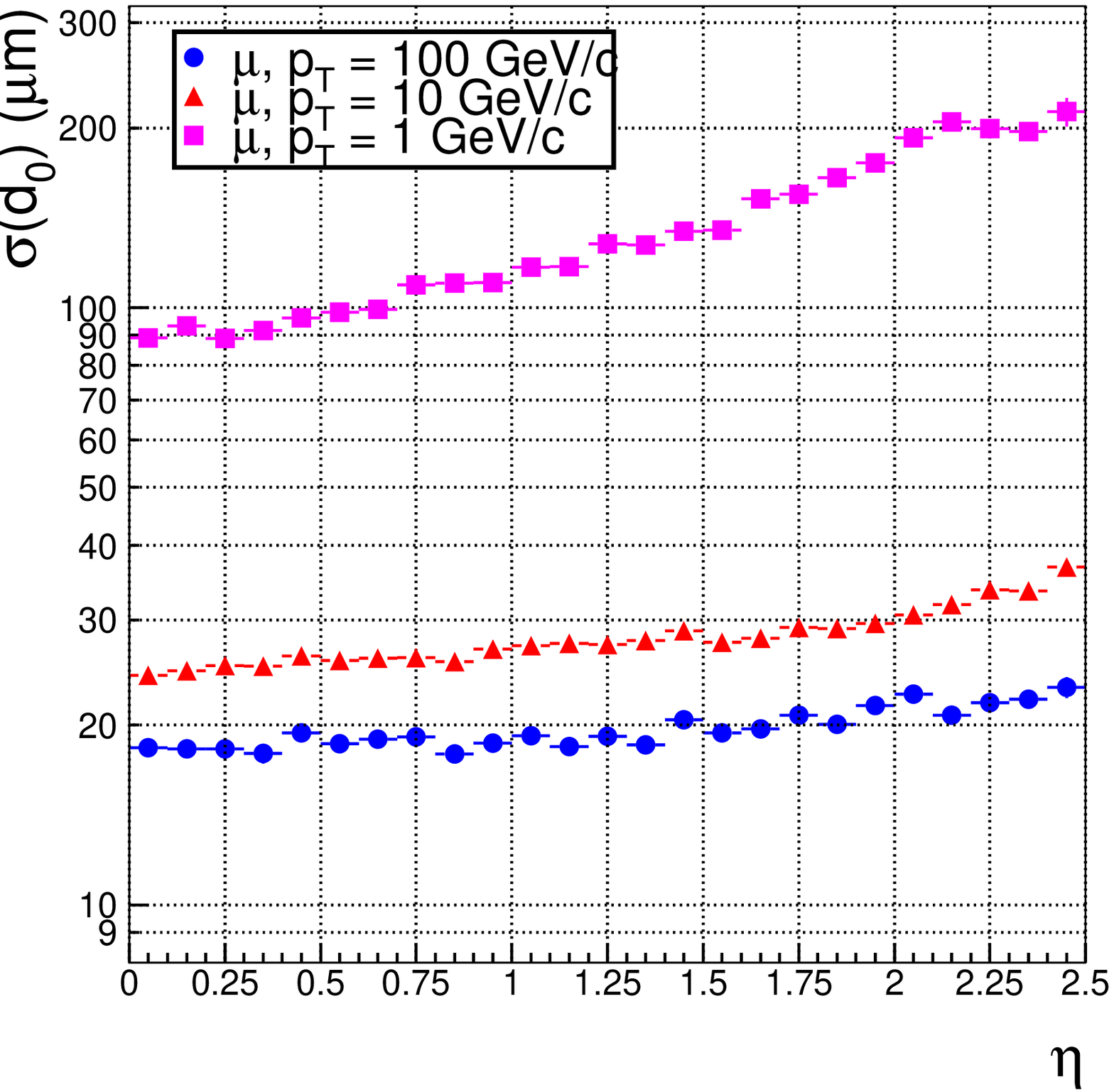}
\caption{\label{fig:perf} 
Performances of the CMS tracker reconstruction of charged particles. 
Left: track reconstruction efficiencies as a function of pseudo-rapidity $|\eta|$
for muons and pions of different tranverse momentum $p_T$.
Center : track transverse momentum resolutions for muons 
as a function of $|\eta|$ and for different  $p_T$ values. 
Right : track impact parameter resolutions for muons 
as a function of $|\eta|$ and for different  $p_T$ values. }
\end{center}
\end{figure}

As it can be seen in figure~\ref{fig:perf} the track reconstruction efficiency
for muons is expected to be close to 100\% in the acceptance range. 
Electrons and hadrons are expected to have slightly worse efficiencies ranging from 
95\% in the central region to 80\% in the forward region.

The transverse momentum resolution for reconstructed tracks is 1-2\% for 100~GeV/c
$p_T$ muons in the central region (figure~\ref{fig:perf}). The resolution is better for lower
momentum muons in the 1-10~GeV/c $p_T$ range, but is degraded in the forward region due 
to the reduced level arm. 
The resolution for the tracks impact parameter is expected to be at the level of 20~$\mu$m
for high energy tracks ($p_T$=100~GeV/c), but degrades at lower momenta 
because of multiple scattering.

The resolution on the impact parameters of reconstructed tracks will be a key factor for 
an efficient identification of final states with b quarks and tau leptons, crucial for 
measurements in the higgs sector with a light higgs decay 
(h$\rightarrow \rm{b}\overline{\rm b}$), for top physics (t$\rightarrow$bW$^+$), and for
searches for new physics like supersymmetric particles. 

\section*{Acknowledgments}
I would like to thank the EPS NPDC19 Conference organizers 
G.~Viesti, A.~Zenoni and A.~Fontana
for kindly inviting me to give a presentation
on the CMS tracker, and for their patience for my delay in sending 
this contributed paper.

\smallskip

\end{document}